# Multiple Field-Induced Phase Transitions in a Geometrically-Frustrated Dipolar Magnet – $Gd_2Ti_2O_7$

A. P. Ramirez[1,a], B. S. Shastry[2,3,5], A. Hayashi[4], J. J. Krajewski[3], D. A.Huse[5], and R. J. Cava[4]

[1]Los Alamos National Laboratory, K764, Los Alamos, NM, 87545
[2]Department of Physics, Indian Institute of Science, Bangalore 560012, India
[3]Bell Laboratories, Lucent Technologies, 600 Mountain Ave., Murray Hill, New Jersey, 07974
[4]Chemistry Department, Princeton University, Princeton, New Jersey, 08540
[5]Physics Department, Princeton University, Princeton, New Jersey, 08540

PACS 75.30.Kz, 75.50.Ee, 75.10.-b

Field-driven phase transitions generally arise from competition between Zeeman energy and exchange or crystal-field anisotropy. Here we present the phase diagram of a frustrated pyrochlore magnet $Gd_2Ti_2O_7$, where crystal field splitting is small compared to the dipolar energy. We find good agreement between zero-temperature critical fields and those obtained from a mean-field model. Here, dipolar interactions couple real-space and spin-space, so the transitions in $Gd_2Ti_2O_7$ arise from field-induced "cooperative anisotropy" reflecting the broken spatial symmetries of the pyrochlore lattice.

The pyrochlore lattice, which consists of corner-shared tetrahedra of spins has special significance in the study of geometrically frustrated systems. The tetrahedron forms a natural basis of a four-sublattice antiferromagnet. However, corner sharing, and the resulting Maxwellian underconstraint, reduce the role of inter-tetrahedral correlations[1]. This balance of local and extended degrees of freedom has led to a variety of unusual collective behaviors, including "cooperative paramagnetism" [2, 3], ultra-small disorder spin glass [4], and spin ice freezing [5, 6]. Of fundamental interest is the interplay between frustrating interactions and applied field. In the Ising spin ice system, $Dy_2Ti_2O_7$, a magnetic field lifts the tetrahedral degeneracy, inducing antiferromagnetic long- range order in a system which was disordered in zero field [6].

In this paper, we present the H-T phase diagram for a dipolar Heisenberg pyrochlore magnet, $Gd_2Ti_2O_7$, as determined by specific heat, C(T), and ac-susceptibility, $\chi$(T). $Gd_2Ti_2O_7$ has a Weiss constant, $\theta_W$, of ~ 10K, [7] and a spin-freezing transition, $T_c$, around 1K, and thus $f \equiv \theta_W/T_c \cong 10$. We find here an unusually complex phase diagram for a Heisenberg system. Agreement between critical field values and mean field theory suggests a novel origin for this complexity – the mapping of broken spatial symmetries of the pyrochlore lattice onto spin space.

Two different types of $Gd_2Ti_2O_7$ samples were studied. Polycrystalline samples of $Gd_2Ti_2O_7$ were synthesized from high purity $Gd_2O_3$ and $TiO_2$ powders, mixed in the stoichiometric ratio, and heated in air in pristine, high-density $Al_2O_3$ crucibles. In all samples chosen for this study, the diffraction peak widths indicated excellent crystallinity, and the refined crystallographic cell parameters were found to be identical within the precision of conventional powder diffraction. Our low-temperature results are different from those previously reported [7]. We therefore carried out experiments probing the possible range of variation in the Gd to Ti ratio and found that the stoichiometric ratio of metals yielded single-phase materials, whereas metal ratio variations of ±5% yielded multiple-phase materials. We view this and also the reproductibility of our results on varying the synthesis route as strong evidence for stoichiometry of the present samples.

For T > 2K, the dc-susceptibility was measured with a commercial magnetometer. Below 1.5 K, ac-susceptibility was measured using a mutual inductance coil technique operating at 143 Hz. For this measurement, the powder was mixed with Stycast 1266 epoxy and cast with an embedded matrix of fine Cu wires for thermal equilibration. The specific heat was measured using a standard semiadiabatic heat-pulse technique. The powder samples were cold-sintered with fine Ag powder to aid thermal equilibration.

In fig. 1 we show $\chi$(T) above 2K. Fitting the data to a straight line above 25 K yields a value for the effective moment of 7.98 $\mu_B$ ($\theta_W$ = -9.4K) in good agreement with the S = 7/2 free-ion value of 7.94 $\mu_B$ and previous work[7]. In the lower inset of fig. 1 we show a C(T) measurement for one of our samples (powder #1). Integrating the data up to 5K yields an entropy of 92% of the full Rln8 expected for $Gd^{3+}$, consistent with the expect small single-ion anisotropy. In fig. 2 we show C(T)/T below 1.5K at H = 0 for three different samples demonstrating the reproducibility of the multiple zero-field transitions. In fig. 2 is also shown C(T)/T for the crystallites sample at a few different field values, demonstrating the merging of the zero-field peaks, followed by splitting at higher fields. Figure 1 (inset) is a surface plot of C(T)/T, showing the full range of data (here, C(T) was measured every 0.5 T up to 8 T). Similar data were obtained for a powder sample.

Ac-susceptibility, $\chi_{ac}$(H), measurements as shown in fig. 3, complement the C(T) data in their greater sensitivity to high-field phase boundaries, and the full phase diagram is shown in fig. 4. Multiple field-induced phase transitions are by themselves not unusual, usually seen in systems with uniaxial anisotropy, and result from from near-degeneracy of phases with different wavevector or symmetry [8]. *The behavior in $Gd_2Ti_2O_7$ is, however, markedly different from other such systems since $Gd^{3+}$ does not typically have significant uniaxial anisotropy.*

Previous zero-field theory on Heisenberg pyrochlores showed that by including long-range superexchange [7], or dipole-dipole coupling [9], an ordered phase is stabilized at low temperature. At zero temperature six distinct ground states are found within a simple 4-sublattice model, and the evolution of these ground states in a field provides features rather similar to our data, as we show below. Since the measurements were made on a polycrystalline sample, the multiple phase transitions as the field is increased may be from crystallites of different orientation. Our model shows either one or two low-temperature phase transitions with increasing field for each of the high-symmetry orientations of the field.

We set up a simple model in the spirit of models used to explain the spin flop transition in uniaxial materials such as $MnF_2$. The model incorporates a dipolar interaction in addition to the superexchange. A mean field treatment of the dipolar interaction [10] yields the phenomenological anisotropy energy. Unlike the case of $MnF_2$, which has uniaxial symmetry in a two-sublattice antiferromagnet, the present case corresponds to cubic symmetry in a four-sublattice magnet. We describe the model and its consequences only qualitatively below, reserving a detailed exposition for a future publication.

In our model, we locate spins on the pyrochlore lattice, viewed as a FCC lattice (primitive vectors a/2(0,1,1), a/2(1,0,1), a/2(1,1,0)), with four basis vectors (0, a/4(0,1,1), a/4(1,0,1), a/4(1,1,0)) defining the four sublattices. We replace each spin on sublattice j by $S\hbar(j)$ and seek to determine the directions of the unit vectors $\hbar(j)$. The problem then reduces to that of just four classical vectors, and the energy is

$$4F/N = +JS^2 \sum_{j,k}{}^{\prime} \hbar(j)\cdot\hbar(k) - gm_B \sum_j \hbar(j)\cdot\vec{B} + (gm_B S)^2/(2a^3) \sum_{i\neq j} \sum_{a,b=x,y,z} \boldsymbol{h}^a(i) D^{ab}(i,j) \boldsymbol{h}^b(j)$$

where g=7/2 and $J$ is the superexchange energy. The dimensionless matrix $D^{ab}$ represents the dipolar energy. This matrix is determined by summing over a sufficiently large spherical region (the Lorenz sphere); evaluation of terms out to a distance of ~$10^2$ sites provides a well-converged result. A version of this problem was considered by Palmer and Chalker [9], who found that in the absence of an external B field this model has six ground states. We find for $B = 0$ precisely the same ground states, and describe these more fully next.

In our treatment, the only adjustable parameter is $J$ and we found fair agreement with the experimental phase boundaries for $J = 0.4$ K. With this value for $J$, and a = 10Å, we find the exchange energy $J \sim 0.60E_d$, where $E_d = 16\sqrt{2}(gm_B S)^2/a^3 \sim 0.653$ K is the dipolar energy for a pair of nearest-neighbor moments. We found the following ground states for B = 0 (in each pattern <ij>, the spins are aligned in the ij plane):

**Pattern <xy>**: $\hbar(1) = \frac{1}{2}\{1,-1,0\}, \hbar(2) = \frac{1}{2}\{-1,-1,0\}, \hbar(3) = \frac{1}{2}\{1,1,0\}, \hbar(4) = \frac{1}{2}\{-1,1,0\}$.

**Pattern <xz>**: $\hbar(1) = \frac{1}{2}\{-1,0,1\}, \hbar(2) = \frac{1}{2}\{1,0,1\}, \hbar(3) = \frac{1}{2}\{1,0,-1\}, \hbar(4) = \frac{1}{2}\{-1,0,-1\}$

**Pattern <yz>**: $\hbar(1) = \frac{1}{2}\{0,1,-1\}, \hbar(2) = \frac{1}{2}\{0,-1,1\}, \hbar(3) = \frac{1}{2}\{0,-1,-1\}, \hbar(4) = \frac{1}{2}\{0,1,1\}$

These patterns and their fully reversed conjugates, e.g. <**xy**>, are evolved in the presence of an external field oriented in various directions and increased in small steps, using standard minimization procedures, we denote the resultants as <**xy**>$_h$ , **<xy>**$_h$ , etc.

In order to study the stability of the mean field states, we compute the frequencies of the antiferromagnetic magnons at zero wave number. We linearize the dynamical equations of motion around a mean field solution and obtain a dynamical matrix for transverse fluctuations. We also compute the Hessian matrix of stability for each mean field solution (fig. 4), by considering variations $\vec{h}(j) \to \vec{h}(j) + \vec{d}(j)$. The vector $\vec{d}$ is orthogonal to $\vec{h}(j)$; the energy is expanded out to quadratic terms in the variation giving the Hessian matrix. Finally, we calculate the (bilinear) tensor "order parameters" $T^{ab} = \frac{1}{4}\sum_{j=1}^{4} \boldsymbol{h}^a(j))\ \boldsymbol{h}^b(j))$ (fig. 4).

The lattice symmetry that is broken when the magnetic field is along <110> is reflection in the plane spanned by the field and the z-axis. The four states $\langle \mathbf{xz} \rangle_h$, $\langle \mathbf{yz} \rangle_h$, $\underline{\langle \mathbf{xz} \rangle}_h$ and $\underline{\langle \mathbf{yz} \rangle}_h$ are energetically degenerate and thermodynamically stable at small field. States $\langle \mathbf{xy} \rangle_h$ and $\underline{\langle \mathbf{xy} \rangle}_h$ have higher energies. The 4 low-field states all have a transition at B ~ 4T, characterized by a deep minimum in the lowest magnon frequency, as well as in the lowest Hessian eigenvalue. At this 4T transition the pairs of states that transform into one another under the reflection operation merge pairwise with $\langle \mathbf{xz} \rangle_h$ and $\langle \mathbf{yz} \rangle_h$ becoming the same state and $\underline{\langle \mathbf{yz} \rangle}_h$ and $\underline{\langle \mathbf{xz} \rangle}_h$ also merging. The two remaining states are degenerate, but not related by any obvious symmetry of the system. Another transition where these two remaining states become the same occurs at B ~ 7 T, again with a vanishing of the lowest magnon and Hessian eigenvalue. Beyond 7T there is only a unique paramagnetic state. Thus both of the transitions are Ising-like, with pairs of states merging.

For field along the <100> direction, the linear magnetization, $M^x$, exhibits no obvious singularities as a function of field. However, there are clear indications of phase transitions in the bilinear order parameters $T^{ij}$. It is perhaps important to point out that similar bilinear operators are the primary order parameters in nematic liquid crystals. As we increase B, the six principal states remain degenerate until a critical field of B ~ 6 Tesla. The relevant symmetry here is rotation about this 4-fold axis. States $<yz>_h$ and $\underline{<yz>}_h$ transform into one another under the symmetry operation consisting of a 90 degree rotation about <100> and a translation. At 6T there is a transition to just two states with the coming together of $<xy>_h$, $<yz>_h$, $<\underline{xy}>_h$ and the other trio of states. This restores the symmetry under 180 degree rotation around <100>. The two remaining states between 6T and 7T transform into one another under the combined 90 degree rotation and translation. At ~7T this symmetry is restored in apparent Ising-like transition.

Note that for the field along <111> there is only the one transition, where both rotation and reflection symmetries are broken simultaneously, and here there are no extra degeneracies that arise from something other than the lattice symmetries.

Thus we see that among the principal directions, a mean field theory reproduces the number of transitions and approximate locations of the three observed critical fields. The transitions occur as the result of the collapse of degeneracies among states that in many, but not all cases, are symmetry-related, as summarized in fig. 5. The surprise here is the high degeneracy that survives in rather large magnetic fields – a situation we believe is uniquely caused by dipolar spins in a geometrically frustrated configuration. Thus we have a situation reminiscent of little groups [11] which form as a result of high symmetry along particular crystallographic directions. Here, however, the high-symmetry results from complex frustrated interactions rather than from a particular crystalline anisotropy.

In summary, we have shown that the geometrically frustrated dipolar Heisenberg magnet $Gd_2Ti_2O_7$ has a complex phase diagram. A minimal mean-field model is presented which produces the low-temperature critical field values. Further experimental work, in particular ESR and polarized neutron scattering, preferably using single crystals, is needed to test our proposed simple model for the various phase transitions.

**Figure Captions**

1. Inverse susceptibility, $1/\chi(T)$, versus temperature for $Gd_2Ti_2O_7$.The dashed line is to guide the eye. The upper inset shows the specific heat divided by temperature, C(T)/T in the full temperature and field range of this work. The lower inset shows the specific heat, C(T), as well as C(T)/T for $Gd_2Ti_2O_7$ over a large temperature scale.

2. Specific heat divided by temperature, C(T)/T, for the crystallites sample of $Gd_2Ti_2O_7$ at different values of applied field. The upper panel also shows C(T)/T at H = 0 for three different samples, two prepared by solid state synthesis and one prepared by arc melting and consisting of large (0.1 mm) crystallites. The powder #1 and powder #2 data are offset by 1.5 and 3.0 J/mole$K^2$ respectively for clarity. The lower panel shows the C(T)/T data in a few finite fields.

3. Ac-susceptibility, $\chi_{ac}$, for $Gd_2Ti_2O_7$ as a function of field at fixed temperatures. The inset shows the full $\chi_{ac}$ dataset for two different temperatures.

4. Upper panel: Phase diagram for $Gd_2Ti_2O_7$. The different symbols denote the position of peaks in either C(T)/T or $\chi_{ac}(H)$. 2nd panel: lowest Hessian eigenvalues as a function of field for two principal directions. 3rd panel: Lowest spin wave energies as a function of field for two principal directions. Bottom panel: The bilinear order parameters $T^{ii}$ for the <xy> state as a function of field along the <100> direction.

5. Evolution and branching of the degenerate phases of the dipolar pyrochlore system as a function of field in the principal directions. Note the vertical direction is not an energy axis.

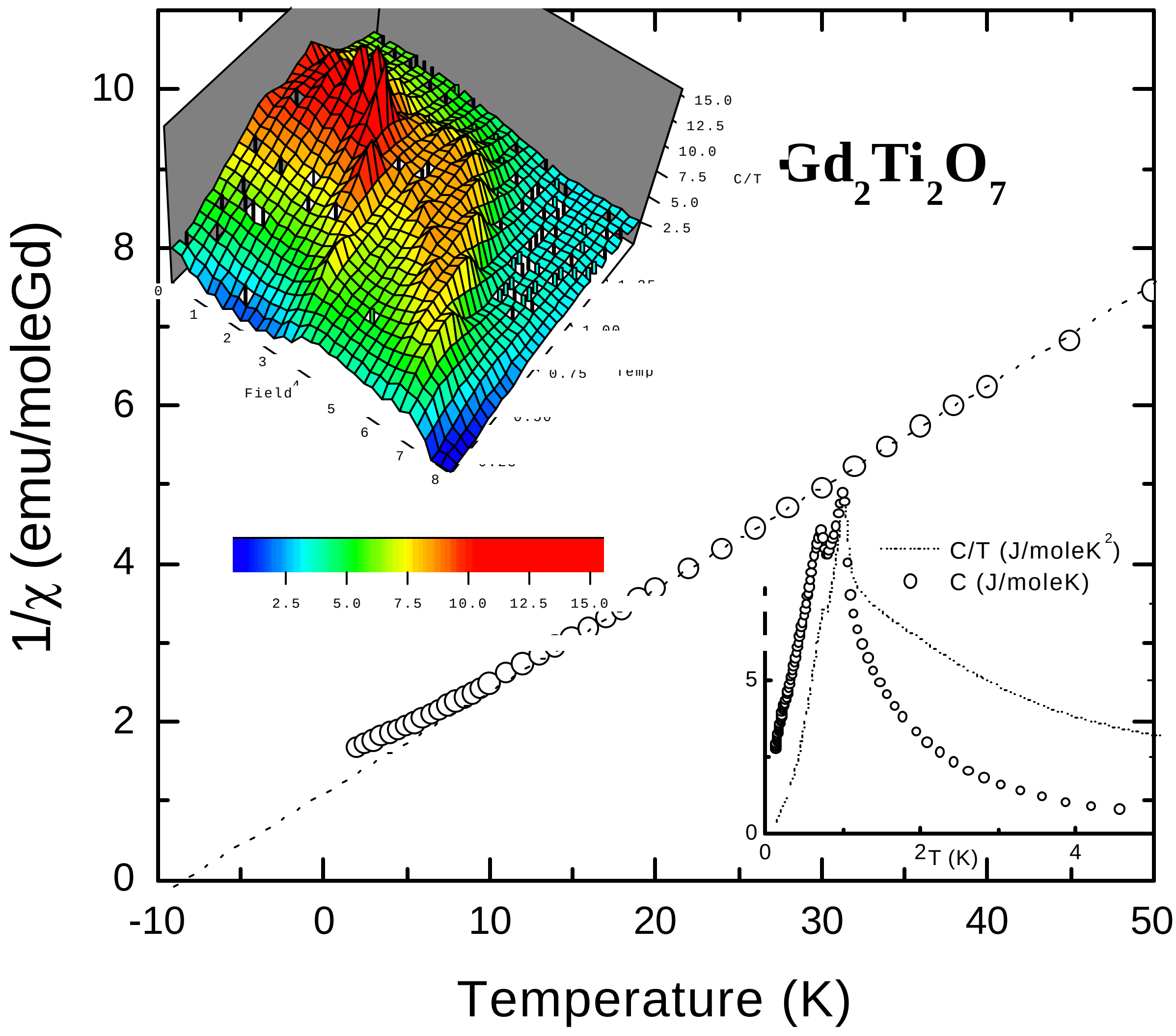

fig. 1

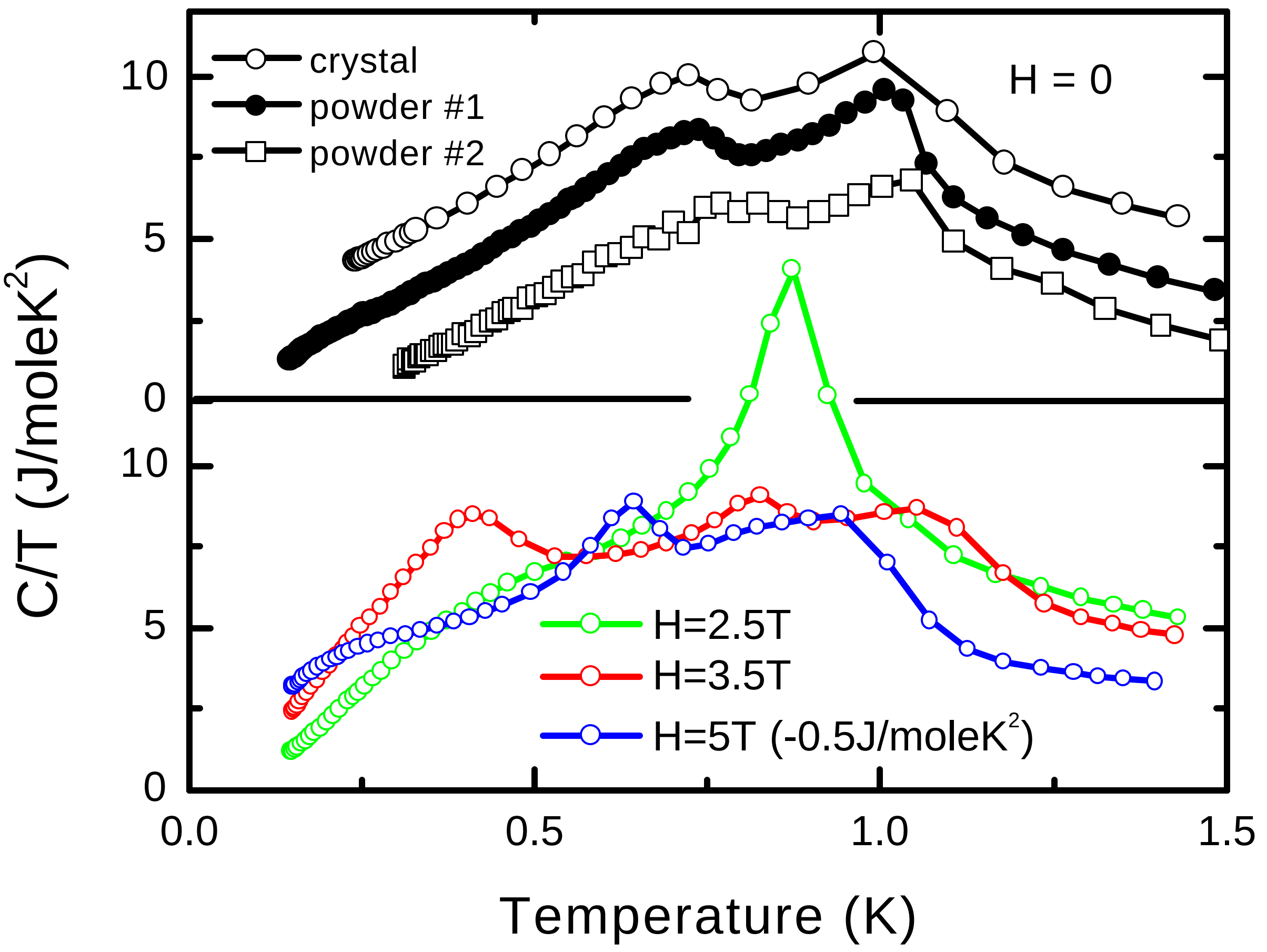

fig. 2

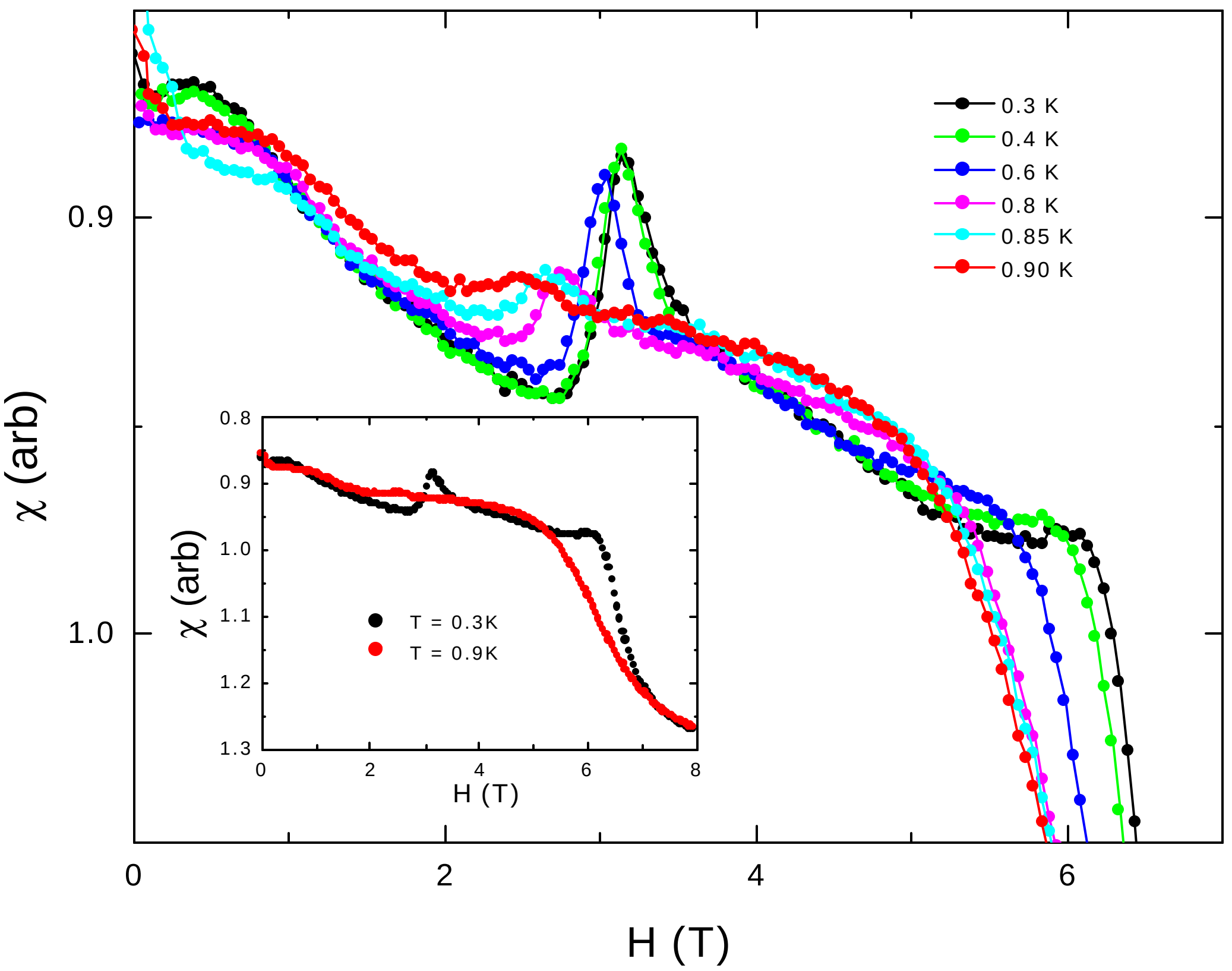

fig. 3

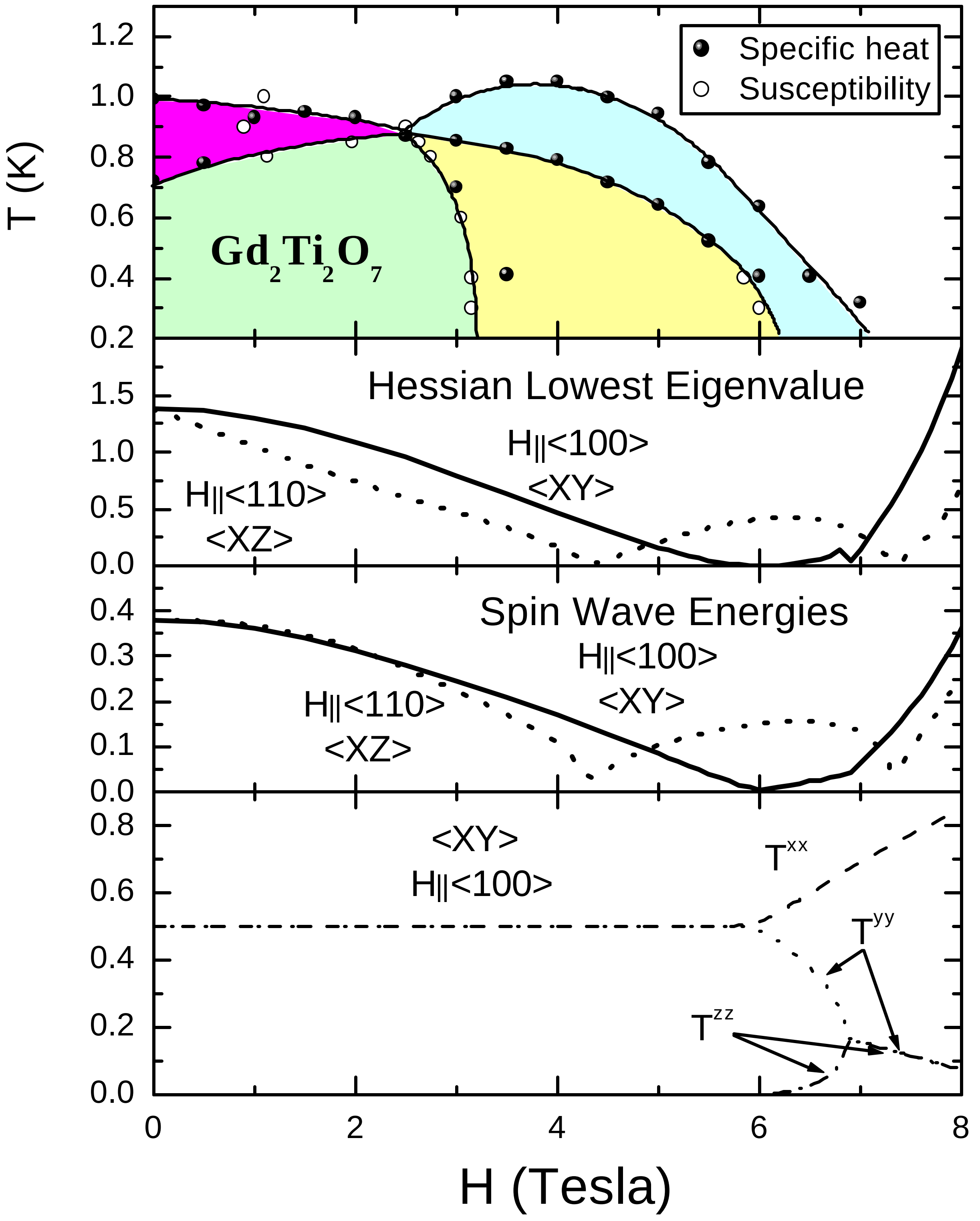

fig. 4

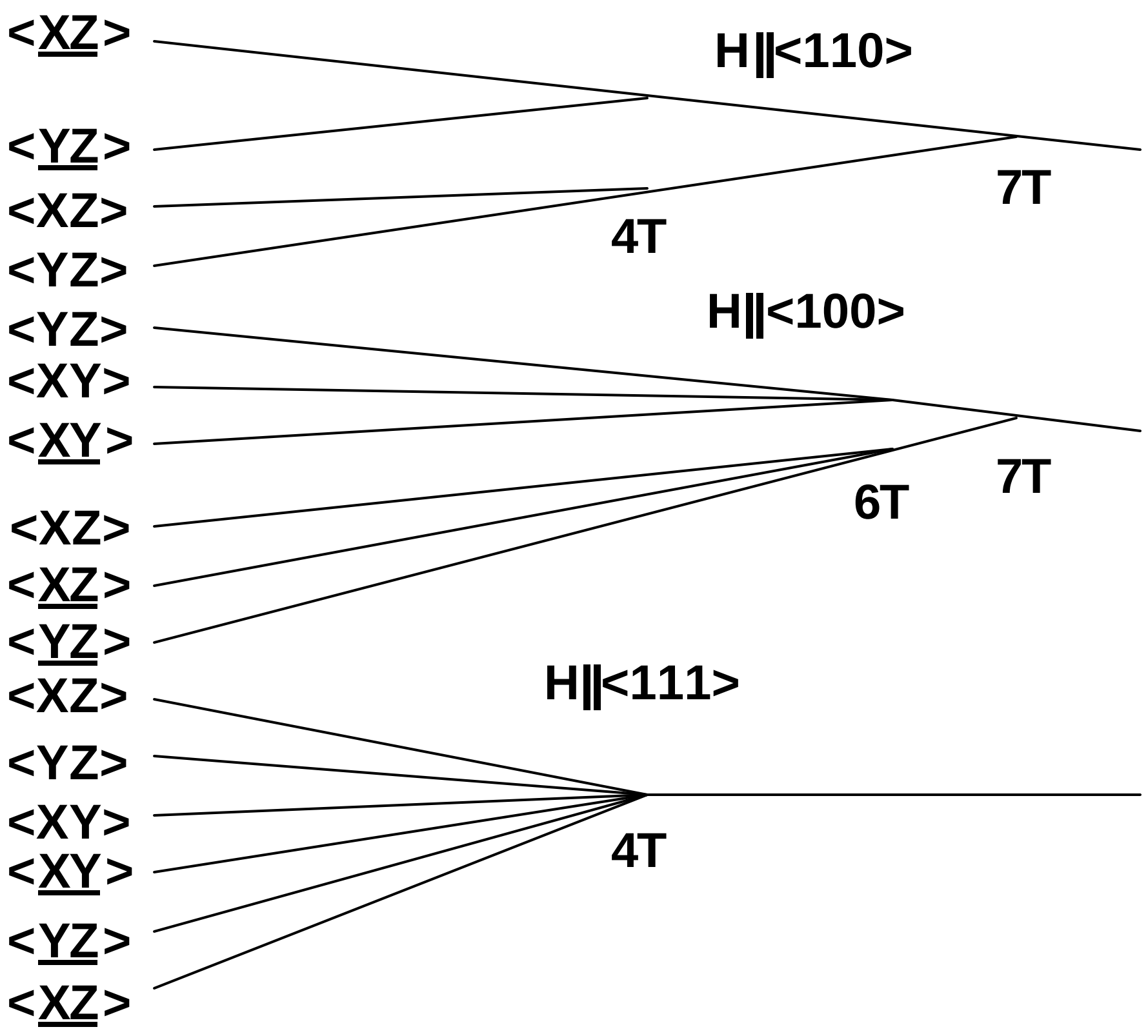

fig. 5